\documentclass[11pt,a4paper]{article}

\usepackage{jcappub}
\usepackage{physics}
\usepackage[capitalise]{cleveref}
\newcommand{\calJ}{\mathcal{J}}
\newcommand{\calP}{\mathcal{P}}
\newcommand{\tstar}{\tau_\star}
\newcommand{\tbr}{\tau_\mathrm{GBR}}

\newcommand{\gbr}{\gamma_\mathrm{GBR}}
\newcommand{\hypergauss}[4]{\phantom{}_{_2}\mathrm{F}\!_{_1}\!\left(#1,#2;#3;#4\right)}

\newcommand{\calF}{n(t,\ell)}
\newcommand{\uHo}{\mathrm{H_{0}}}
\newcommand{\zf}{z_{\mathrm{friction}}}

\newcommand{\odgrb}{\omega_\mathrm{DGRB}}
\newcommand{\OmegaR}{\Omega_\mathrm{rad}}
\newcommand{\OmegaM}{\Omega_\mathrm{mat}}

\title{A window for cosmic strings}

\arxivnumber{2112.11093}

\usepackage{soul}

\author[a,b]{Pierre Auclair}
\author[a]{Konstantin Leyde}
\author[a]{Dani\`ele A.~Steer}

\affiliation[a]{Universit\'e Paris Cit\'e, CNRS, Astroparticule et Cosmologie, F-75013 Paris, France}
\affiliation[b]{Cosmology, Universe and Relativity at Louvain, Institute of Mathematics and Physics, Louvain University, 2 Chemin du Cyclotron, 1348 Louvain-la-Neuve, Belgium}

\emailAdd{pierre.auclair@uclouvain.be}
\emailAdd{kleyde@apc.univ-paris7.fr}
\emailAdd{steer@apc.univ-paris7.fr}

\abstract{
	Particle emission, in addition to gravitational radiation from cosmic string loops, affects the resulting loop distribution and hence the corresponding observational consequences of cosmic strings.
	Here we focus on two models in which loops of length $\ell$ are produced from the infinite string network with a given power-law.
	For both models we find that, due to particle production, the Stochastic Gravitational Wave Background (SGWB) is cut off outside the region of parameter space probed by any current or planned GW experiment.
	Therefore the present constraints from the LIGO-Virgo-Kagra (LVK) collaboration still hold.
	However for one of these models, if a fraction $\gtrsim \order{10^{-3}}$ of these particles cascades into $\gamma$-rays, and if the gravitational backreaction scale follows the Polchinski-Rocha model, then the string tension is tightly constrained from below by measurements of the Diffuse $\gamma$-Ray Background, and from above by the SGWB.
	With reasonable assumptions, the joint constraint on the string tension set by these two possible observables reduces the available parameter space of this cosmic string model to a narrow  band. Future upgrades to LVK will either rule out this model or detect strings.
}

\begin{document}

\maketitle

\tableofcontents

\section{Introduction}

Cosmic strings are line-like topological defects which may form in symmetry breaking phase transitions, provided the vacuum manifold contains non-contractible loops \cite{Kibble:1976sj,Hindmarsh:1994re,Vilenkin:2000jqa, Vachaspati:2015cma}.  Due to their topological stability, any strings formed in the early universe will be present throughout the history of the universe, and thus can leave
observational consequences which may be visible today. These include particle emission from strings (observed for example as high energy cosmic rays or a Diffuse $\gamma$-Ray Background), lensing of galaxies, CMB fluctuations generated by strings, gravitational wave (GW) emission from strings in the form of short bursts or a stochastic GW background (SGWB) (see Ref.~\cite{Hindmarsh:1994re,Vilenkin:2000jqa,Vachaspati:2015cma} for reviews).
If detected, cosmic strings can thus probe the corresponding energy scale $\eta$ of the symmetry breaking phase transition during which they were formed.

In this paper, we extend a previous publication by some of the authors~\cite{Auclair:2019jip}, and focus on the \emph{combined} constraints from GWs at LIGO-Virgo-Kagra frequencies (as well as predictions for LISA) and the diffuse $\gamma$-ray background through FERMI-Lat.  The novel aspect of this work is to consider models in which cosmic string loops of all sizes are produced from the infinite string network with a given power-law.  Depending on the properties of the power-law --- and in particular for  the Polchinski-Rocha model --- we show that these two observations already constrain the string parameters very strongly and essentially close the window on the allowed parameter space to a very small region. This will be further reduced or excluded through future GW observations, or strings will be detected.

We consider local, non current-carrying cosmic strings, parametrized by the dimensionless string tension, $G\mu$ (where $G$ is Newton's constant) where
\begin{equation}
	G\mu \sim 10^{-6} \qty(\frac{\eta}{10^{16} ~\mathrm{GeV}})^2.
\end{equation}
The corresponding microscopic string width, $w$, is given by
\begin{equation}
	w\sim \mu^{-1/2} \sim 1/\eta,
	\label{wdef}
\end{equation}
and is much smaller than characteristic macroscopic string size $\ell \gg w$ \cite{Vachaspati:2015cma}.
Using the Nambu-Goto equations of motion and the `intercommutation' of strings\footnote{The standard assumption, shown to be valid in for the collision of local $U(1)$ strings is that the strings `exchange partners' or `intercommute'~\cite{Shellard:1987bv,Matzner:1988qqj} (there are some exceptions for high velocity collisions \cite{Verbiest:2011kv}).}, the evolution of a network of strings formed at energy scale $\eta$ can then be studied. Intercommutation leads to the formation of closed loops of string which lose energy through GWs or other particle radiation, thus extracting energy from the remaining long-string network which reaches a scaling solution, see e.g.~\cite{Vilenkin:2000jqa}. When gravitational radiation is the dominant energy decay mechanism, the average power in GWs emitted from a loop of length $\ell$ given by \cite{Vachaspati:1984gt,Burden:1985md,Garfinkle:1987yw,Hindmarsh:1990xi,Allen:1991bk,Damour:2001bk,Siemens:2001dx,Siemens:2006vk,Blanco-Pillado:2017oxo}
\begin{equation}
	P_\mathrm{GW} = \Gamma G \mu^2\,,
	\label{eq:pgw}
\end{equation}
where the constant $\Gamma$  generally depends on the specific shape of the loop of length $\ell$, but on average $\Gamma \simeq 50$ for all $\ell$. The integrated effect of all the GWs emitted by numerous loops of string formed during the evolution of the network from string formation until today, is a stochastic GW background (SGWB).

The number density $n(t,\ell)$ of closed loops of length $\ell$ at time $t$ in a string network depends on the loop production function $\calP(t, \ell)$, which describes the sourcing of loops by intercommutation.
Its precise form depends on complex underlying physics, including gravitational backreaction (GBR) effects which can be important at points of high curvature (see \cite{Blanco-Pillado:2019nto} for studies of individual loops). The situation is further complicated by the non-linear dynamics of the network and the large range of scales in the problem --- from the horizon size $H^{-1}$ to the scale of gravitational radiation $(\Gamma G\mu) t$ and down to the string width $w$. Numerical studies of Nambu-Goto strings do not include GBR, but can be used to infer $\calP(t, \ell)$ on scales $\ell \gtrsim \Gamma G\mu$, while on smaller scales this can be extended with analytical work. In this paper we focus on power-law loop production functions of the form
\begin{equation}
	\label{eq: def polchinski rocha production}
	t^5 \calP(t,\ell) = C \qty(\frac{\ell}{t})^{2\chi-3} \Theta\qty(\gamma_\infty - \frac{\ell}{t}) \Theta\qty(\frac{\ell}{t} - \gbr)\,,
\end{equation}
where $\chi$ is related to the fractal dimension of the infinite strings.

Using numerical simulations, in \cite{Blanco-Pillado:2013qja}, $\calP(t,\ell)$ was shown to be of the form \eqref{eq: def polchinski rocha production} in both the radiation and matter eras, with parameters ($\gamma_\infty \sim 0.1,C,\chi$) determined from the simulation (see \cref{tab:numirical values}), and the GBR scale assumed to be $\gbr \simeq \Gamma G\mu$.
For those parameters, and as concerns the resulting loop distribution $n(t,\ell)$, it was shown \cite{Blanco-Pillado:2013qja} that $\calP(t,\ell)$ can be effectively be replaced by $\calP(t,\ell)\propto \delta(\ell-\gamma_\infty t)$.
Thus all loops are effectively formed at the same size $\gamma_\infty t$ (see also Refs.\cite{Auclair:2019wcv,Abbott:2021ksc} for more details).
This $\delta$-function approximation is known as Model A in the LVK publications~\cite{Abbott:2021ksc} and was analysed in our previous paper \cite{Auclair:2020wse}.

Other authors have studied loop production analytically, including the effects of GBR. References \cite{Polchinski:2006ee,Polchinski:2007qc,Dubath:2007mf} (see also \cite{Copeland:2009dk} for work on small scale structure and kinks on strings) also obtain a power-law loop production function of the form \eqref{eq: def polchinski rocha production}, where now  \cite{Polchinski:2007rg}
\begin{equation}
	\gbr = \Upsilon (G\mu)^{1+2\chi}
	\label{eq:upsilon}
\end{equation}
with $\Upsilon = \order{20}$. In order to obtain a loop distribution compatible with the simulations \cite{Ringeval:2005kr} on scales $\ell \gtrsim \Gamma G\mu$, the parameters $(C,\chi)$ now take different values --- these are given in \cref{tab:numirical values}.
This so-called Model B, for which $\gbr \ll \Gamma G \mu$, was also used by the LVK collaboration \cite{Abbott:2021ksc}. It has a similar loop distribution to Model A on large scales, but a very different loop distribution on smaller scales \cite{Lorenz:2010sm,Auclair:2019zoz}.

Other than being tested by the LVK-collaboration, both Models A and B have been studied by the LISA consortium \cite{Auclair:2019wcv,LISACosmologyWorkingGroup:2022jok}, as well as considered in numerous other publications.
However, as was first mentioned in the early work of Polchinski and collaborators \cite{Polchinski:2007qc}, see also Ref.~\cite{Blanco-Pillado:2019vcs}, understanding energy conservation (between the long string network and the loop distribution) is not straightforward and is potentially problematic for Model B.
In Appendix \cref{app:energy-con} we discuss this point in more detail, and explain how including the important physical effect of loop fragmentation (which has been observed in multiple simulations but is not included in Ref. \cite{Blanco-Pillado:2019vcs}), modifies these energy conservation arguments.
This in particular means that the situation regarding Model B remains open in our opinion.
We thus follow other publications, and take Model B as given.
The GW constraints on cosmic strings from the LVK collaboration on the two models are~\cite{Abbott:2021ksc}:
\begin{align}
	{\text{Model A}}: \qquad G\mu & \lesssim 9.6 \times 10^{-9}  \\
	{\text{Model B}}: \qquad G\mu & \lesssim 4.0 \times 10^{-15}
\end{align}
The forecasts from LISA are that the SGWB will be able to constrain $G\mu$ down to $10^{-17}$, for both models A and B \cite{Auclair:2019wcv}.

A different approach to studying the evolution of a cosmic string network is to solve the underlying field equations, rather than the NG equations of motion, using high resolution field theory simulations. Despite the huge range of scales in the problem, this approach has been taken in a series of papers \cite{Vincent:1997cx,Hindmarsh:2008dw,Lizarraga:2016onn,Hindmarsh:2017qff}.  Scaling of the infinite string network is observed as in NG simulations, but in these works the majority of loops are seen to decay directly into particle radiation. As a result there are few long-lived non self-intersecting loops, meaning that these field theory simulations suggest that cosmic strings would on the contrary source a negligible SGWB.  As the NG approach and the field theory approach are in principle describing the same physics, this ongoing situation still requires a resolution.

A step was taken in this direction by Matsunami et al.\,\cite{Matsunami:2019fss} who performed high resolution field theory simulations of individual loops of length $\ell$ containing \emph{kinks}. They identified a new characteristic length scale $\ell_0$, and argued that loops greater than this scale decay into GWs, whereas smaller loops decay into particles. We model this by 
\begin{equation}
	\dv{\ell}{t} = \begin{cases}
		-\Gamma G\mu ,                                          & \ell \gg \ell_{0}  \\
		-\Gamma G\mu \left( \dfrac{\ell_{0}}{ \ell} \right)^n , & \ell \ll \ell_{0},
	\end{cases}
	\label{kink}
\end{equation}
so that once a loop starts decaying into particles its remaining lifetime scales as $\ell^{n+1}$. For loops with kinks $n=1$, and the scale $\ell_0$ is determined by the string width $w$, \cref{wdef}, through
\begin{equation}
	\ell_{0, \mathrm{kink}} = \beta_\mathrm{kink} \frac{w}{\Gamma G\mu},
	\label{eq:ell0-kink}
\end{equation}
with $\beta_\mathrm{kink} = \order{1}$. 
We note that the quadratic lifetime of such kinky loops has been debated in the subsequent analysis of Ref.~\cite{Hindmarsh:2021mnl}, who argue that ``naturally produced'' loops decay rapidly into particles with a linear lifetime ($n=0$). In other words, they argue that the behaviour found in Ref.~\cite{Matsunami:2019fss} may be an artefact of the loop formation method used there.
In the following, we use the decay rate  \cref{kink} which is suitable to describe the loops with a quadratic lifetime of Ref.~\cite{Matsunami:2019fss}. The loops with linear lifetime of Ref.~\cite{Hindmarsh:2021mnl} would require another parametrization, closer to that which has been proposed to study vortons, e.g.~$\dv*{\ell}{t} = \Gamma_1 G\mu \Theta(\ell - \ell_0) + \Gamma_2 G\mu \Theta(\ell_0 - \ell)$~\cite{Peter:2013jj}.\footnote{One could consider a model with a fraction of loops having a quadratic lifetime, and the remainder linear, but given the uncertainties in the literature we do not follow this route.}

For loops with cusps, the situation is more consensual.
Previous estimates by Refs.~\cite{Blanco-Pillado:1998tyu, Olum:1998ag} have shown that these are modeled by taking $n=1/2$: we reproduce this computation analytically using the Kibble-Turok family of solutions~\cite{Kibble:1982cb} in \cref{sec:cusp_analytic}. Furthermore
\begin{equation}
	\ell_{0, \mathrm{cusp}} = \beta_\mathrm{cusp} \frac{w}{(\Gamma G\mu)^2},
	\label{eq:ell0-cusp}
\end{equation}
with $\beta_\mathrm{cusp} = \order{1}$.

In a previous paper \cite{Auclair:2019jip}, we have calculated the effect of this particle production on Model A under the simplifying assumption that all the loops are produced with the same length.
Our aim in this paper is to carry out a similar analysis but for the power-law loop production function of \cref{eq: def polchinski rocha production}.
The calculation is much more involved and enables us also to probe both Models A and B.
In \cref{sec:lnd} we calculate the effect of an $\ell$-dependent energy loss
\begin{equation}
	\dv{\ell}{t} = - \Gamma G\mu \calJ(\ell),
	\label{eq:new}
\end{equation}
on the loop distribution $n(t, \ell)$.
First, in \cref{sec:continuity} we present the general continuity equation satisfied by $n(t, \ell)$, then in \cref{sec:polchinski} we solve it for a power-law loop production function as given in \cref{eq: def polchinski rocha production}.
The observational signatures of the resulting loop distribution are then calculated in \cref{sec:obs}: first we focus on the SGWB and then on the predicted diffuse gamma ray flux. Then, in \cref{sec:joint}, we combine these constraints and determine the allowed regime of $G\mu$ for both model A and B.
Finally, we conclude in \cref{sec:conc}.

\section{The loop distribution with particle production}\label{sec:lnd}

Our aim in this section is to determine  the distribution of non self-intersecting loops $\calF$ taking into account the energy lost in terms of particles. That is $\dv*{\ell}{t}$ is given by \cref{eq:new} with
\begin{equation}
	\label{eq:loop decay}
	%\dv{\ell}{t} = - \Gamma G\mu \calJ(\ell) ,\quad
	\calJ(\ell) \equiv \Theta(\ell - \ell_0) + \frac{\ell_0^n}{\ell^n} \Theta(\ell_0 - \ell),
\end{equation}
where $n\geq 0$.
In \cref{sec:continuity} we recall the continuity equation satisfied by $\calF$ (see also \cite{Copeland:1998na}) and the procedure to solve it using a particular change of variable and a Green's function.
Then, in \cref{sec:polchinski}, we solve this equation for the power-law loop production function of \cref{eq: def polchinski rocha production}. %and illustrate its different sub-populations in \cref{fig:loop production and ell0}

\subsection{Continuity equation for non self-intersecting loops}
\label{sec:continuity}

The distribution of non self-intersecting loops is assumed to satisfy the continuity equation \cite{Copeland:1998na, Peter:2013jj, Auclair:2019jip}
\begin{equation}
	\label{eq: continuity equation}
	\eval{\pdv{t}}_{\ell} \qty[a^3 \calF] + \eval{\pdv{\ell}}_{t}\qty[\dv{\ell}{t} a^3 \calF] = a^3 \calP(t,\ell)\,,
\end{equation}
where $a(t)$ is the scale factor, and the loop production function $\calP(t,\ell)\dd{\ell}\dd{t}$ quantifies the number of non self-intersecting loops with invariant length $\in [\ell,\ell+\dd{\ell}]$ produced at times $\in [t,t+\dd{t}]$. The link between the continuity equation and the Boltzmann equation is discussed in Ref.~\cite{Auclair:2020wse}.

In a previous paper \cite{Auclair:2019jip} we showed that the general solution to \cref{eq: continuity equation} is much simplified by the introduction of the following coordinates
\begin{align}
	\label{def: xi}
	\hat\tau(t)   & \equiv \Gamma G \mu t                        \\
	\label{eq: def xi}
	\hat\xi(\ell) & \equiv \int \dfrac{\dd{\ell}}{\calJ(\ell)} =
	\begin{cases}
		\dfrac{\ell^{n+1}}{(n+1) \ell_0^n}\, , & \ell < \ell_0     \\[2ex]
		\ell - n \dfrac{\ell_0}{(n+1)}\, ,     & \ell > \ell_0\, ,
	\end{cases}
\end{align}
where $\hat \xi$ is defined only up to an additive constant, that we set so that $\hat\xi(0) = 0$.
In terms of these coordinates a string of length $\ell$ created at a time $t$, has the constant of motion $2v$, given by
\begin{equation}
	2v \equiv \hat\tau(t) + \hat\xi(\ell)\,.
	\label{eq:v}
\end{equation}
The continuity equation then reduces to a wave equation, with general solution
\begin{equation}
	\label{eq: n solution general }
	n( t, \ell) = \frac{1}{\Gamma G\mu \calJ( \ell) a^3( t)} \int \dd{\tau'} a^3(\tau') \calJ\qty[\hat\ell\qty(\xi + \tau - \tau' )] \calP\qty[\tau', \hat\ell\qty(\xi + \tau - \tau' )]\,,
\end{equation}
where the function $\hat\ell$ is given by
\begin{equation}
	\hat\ell(\xi) =
	\begin{cases}
		\qty[(n+1) \ell_0^n \xi]^{1/(n+1)} & \ell < \ell_0 \\
		\xi + n\dfrac{\ell_0}{n+1}         & \ell > \ell_0
	\end{cases}\,,
\end{equation}
and we have set $ \tau = \hat\tau( t)$ and $\xi = \hat\xi({\ell})$.

\begin{figure}
	\centering
	\includegraphics[width=.49\textwidth]{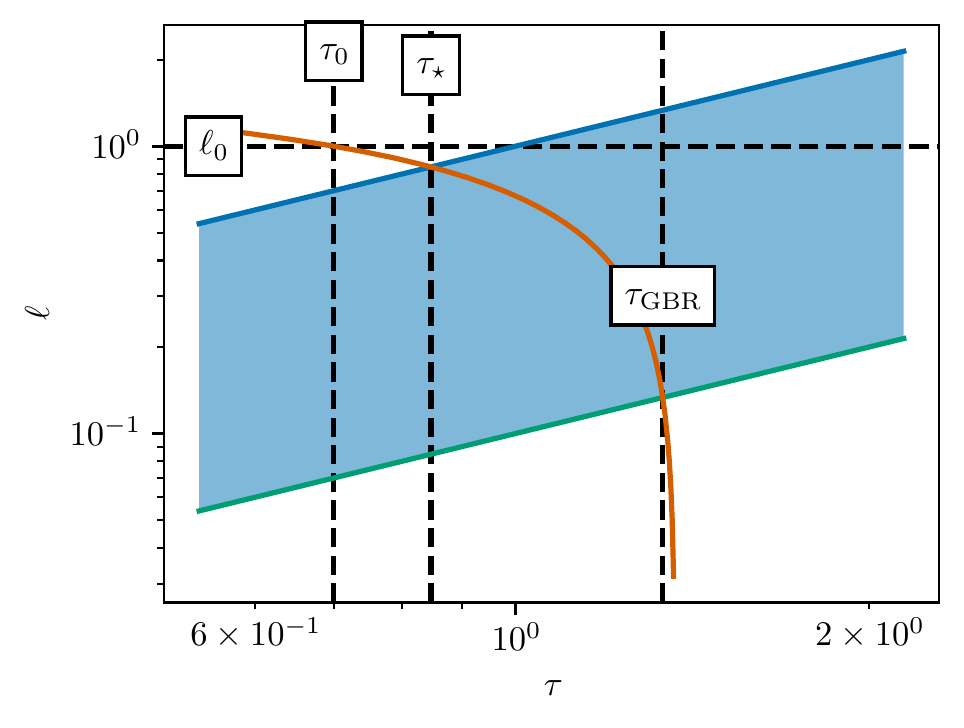}
	\includegraphics[width=.49\textwidth]{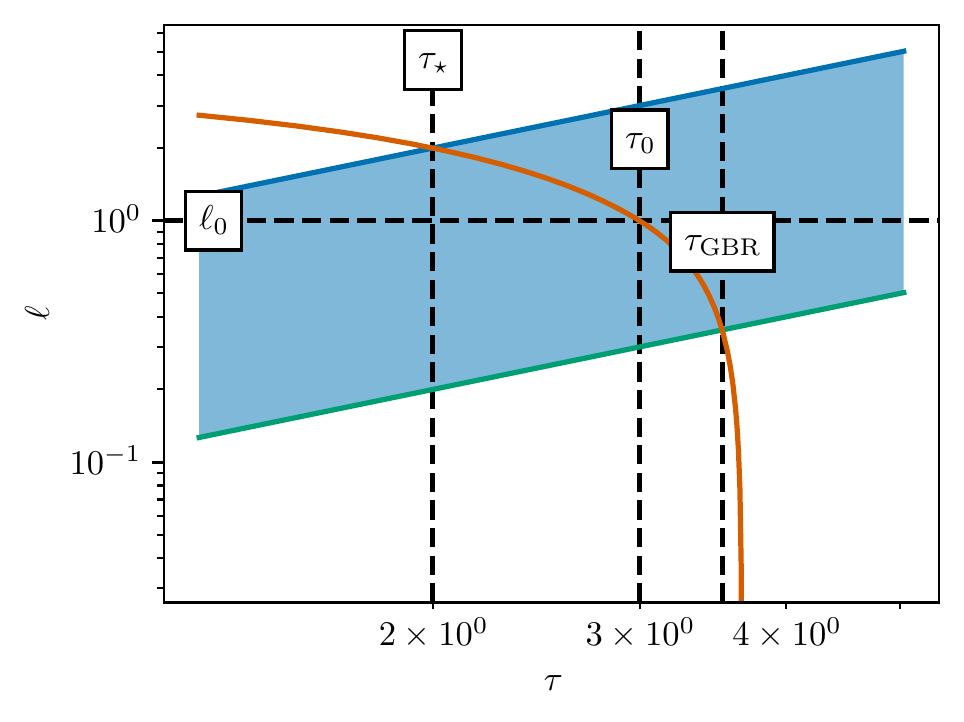}
	\includegraphics[width=.49\textwidth]{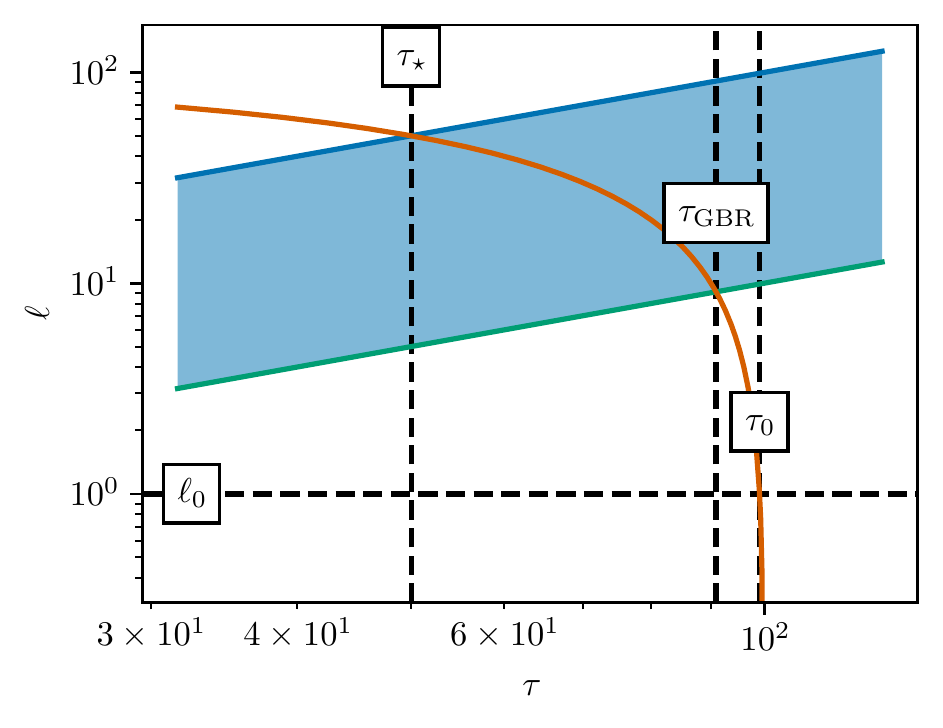}
	\caption{Schematic trajectories of non self-intersecting loops in the $(\tau,\ell)$ phase space (orange). The blue region corresponds to the production of loops, that is the green line is $\gbr t$, and the blue one is $\gamma_\infty t$. These delimit the $\Theta$-function shown in the loop production function of \cref{eq: def polchinski rocha production}.
	There are three distinct possibilities for the ordering of $\tstar$, $\tau_0$ and $\tbr$, which appear in the boundaries of \cref{eq: final result n}.
	This depends on whether the loop had length $\ell_0$ before (\textit{top left}), during (\textit{top right}) or after (\textit{bottom}) loop production.}
	\label{fig:loop production and ell0}
\end{figure}

\subsection{Solution for a power-law loop production function}
\label{sec:polchinski}

We now choose the power-law loop production function \cref{eq: def polchinski rocha production}, and obtain explicit solutions to \cref{eq: n solution general }.

The overall constant $C$ in \cref{eq: def polchinski rocha production} depends on whether strings are produced in the matter or radiation dominated era.
Thus, we denote it by $C_\nu$, where $\nu=1/2$ and $\nu=2/3$ label the radiation and matter era, respectively.
The numerical values for all the parameters for Models A and B are given in \cref{tab:numirical values}.
For purely Nambu-Goto strings, \emph{i.e.~}$\forall \ell, \, \calJ(\ell) = 1$, the resulting loop distribution has already been calculated in Ref.~\cite{Lorenz:2010sm}.
The ultraviolet cutoff $\gbr$, set by the gravitational backreaction scale, prevents the divergence of the loop distribution on small scales. It is given in \cref{eq:upsilon}.
The IR cutoff $\gamma_\infty$, of the order of the Hubble horizon, is necessary to obtain scaling solutions in some regions of the $(\chi_\mathrm{rad}, \chi_\mathrm{mat})$ parameter space (see Ref.~\cite{Auclair:2019zoz} for more details).

\begin{table}
	\begin{center}
		\centering
		\begin{tabular}{|c|c|c|}
			\hline
			Parameter & Model A & Model B \\
			\hline
			$C_\mathrm{rad}$    & $0.28$  & $8\times 10^{-3}$    \\
			$C_\mathrm{mat}$    & $0.17$  & $6.15\times 10^{-3}$ \\
			$\chi_\mathrm{rad}$ & $0.5$   & $0.2$                \\
			$\chi_\mathrm{mat}$ & $0.695$ & $0.295$              \\
			$\gbr$              & $\Gamma G \mu$ & $\Upsilon (G\mu)^{1+2\chi}$ \\
			$\gamma_\infty$     & $0.1$   & $0.1$                \\
			\hline
		\end{tabular}
		\caption{With the exception of $\gbr$, the above values were calibrated using the results of the simulations~\cite{Blanco-Pillado:2013qja} for Model A, and the simulations~\cite{Ringeval:2010ca} for Model B. }
		\label{tab:numirical values}
	\end{center}
\end{table}

Substitution of \cref{eq: def polchinski rocha production} into \cref{eq: n solution general } gives
\begin{equation}
	\label{eq: n for PolchinskiRocha}
	n( t, \ell) = \frac{C_\nu \alpha_{\nu}^3}{(\Gamma G\mu)^{\epsilon} \calJ( \ell) a^3( t)} \int_{\min(\tstar,\tau)}^{\min( \tau, \tbr)} \dd{\tau'} \tau'^{\epsilon-1} \calJ\qty[\hat\ell\qty(\xi + \tau - \tau' )] \hat\ell^{2\chi - 3}\qty(\xi + \tau - \tau' )\,,
\end{equation}
where $\epsilon\equiv 3\nu-2\chi-1$, and at the time $t'$ of formation of the loop, we have approximated the scale factor by $a(t') = \alpha_\nu t'^\nu$. (This is equivalent to assuming an instantaneous radiation-matter transition, and ensures that the integral is analytically tractable.)
The boundaries of the integral are the following: $\tstar$ is the earliest possible time for a loop to have been formed, and $\tbr$ is the latest.
Their explicit expressions are given in \cref{app: calculation of tau technical}.
Due to the choice of the loop decay function in \cref{eq:loop decay},
the above integral splits into two parts, delimited by the time $\tau_0$ when the loop had length $\ell_0$:
\begin{equation}
	\label{eq: tau0}
	\tau_0 = \xi + \tau - \frac{\ell_0}{n+1}\,.
\end{equation}
\cref{fig:loop production and ell0} shows the evolution of the length of the loop as a function of time, and the possible orderings for the timescales $\tstar, \tbr$ and $\tau_0$.
To summarize, we obtain the following for the loop number density
\begin{multline}
	\label{eq: final result n}
	n( t,\ell) = \\ \frac{(\Gamma G\mu)^{-\epsilon} C_\nu \alpha_{\nu}^3}{\epsilon \calJ(\ell) a^3( t) }\left\{\ell_0^{\frac{2n(\chi-1)}{n+1}}
	\qty[(n+1)2 v]^{\frac{2\chi-3-n}{n+1}}
	\left[\tau'^\epsilon\hypergauss{\frac{-2\chi+3+n}{1+n}}{\epsilon}{\epsilon+1}{\frac{\tau'}{2 v}}\right]_{\max(\tstar, \tau_0)}^{\min(\tau,\tau_{\mathrm{GRB}})} \right. \\
	+
	\left.
	\left(2 v+\ell_0\frac{n}{n+1}\right)^{2\chi-3}\left[\tau'^\epsilon\hypergauss{-2\chi+3}{\epsilon}{\epsilon+1}{\frac{\tau'}{2 v+\ell_0\frac{n}{n+1}}}\right]_{\tstar}^{\min(\tau, \tbr, \tau_0)}\right\}\,,
\end{multline}
where $2 v \equiv \hat\xi(\ell) +  \hat\tau(t)$, and the hypergeometric function satifies
\begin{equation}
	\int x^{\beta-1} (1-x z)^{-\alpha} \dd{x} = \frac{x^\beta}{\beta} \hypergauss{\alpha}{\beta}{\beta + 1}{x z}\,.
\end{equation}

\begin{figure}
	\centering
	\includegraphics[width=.49\textwidth]{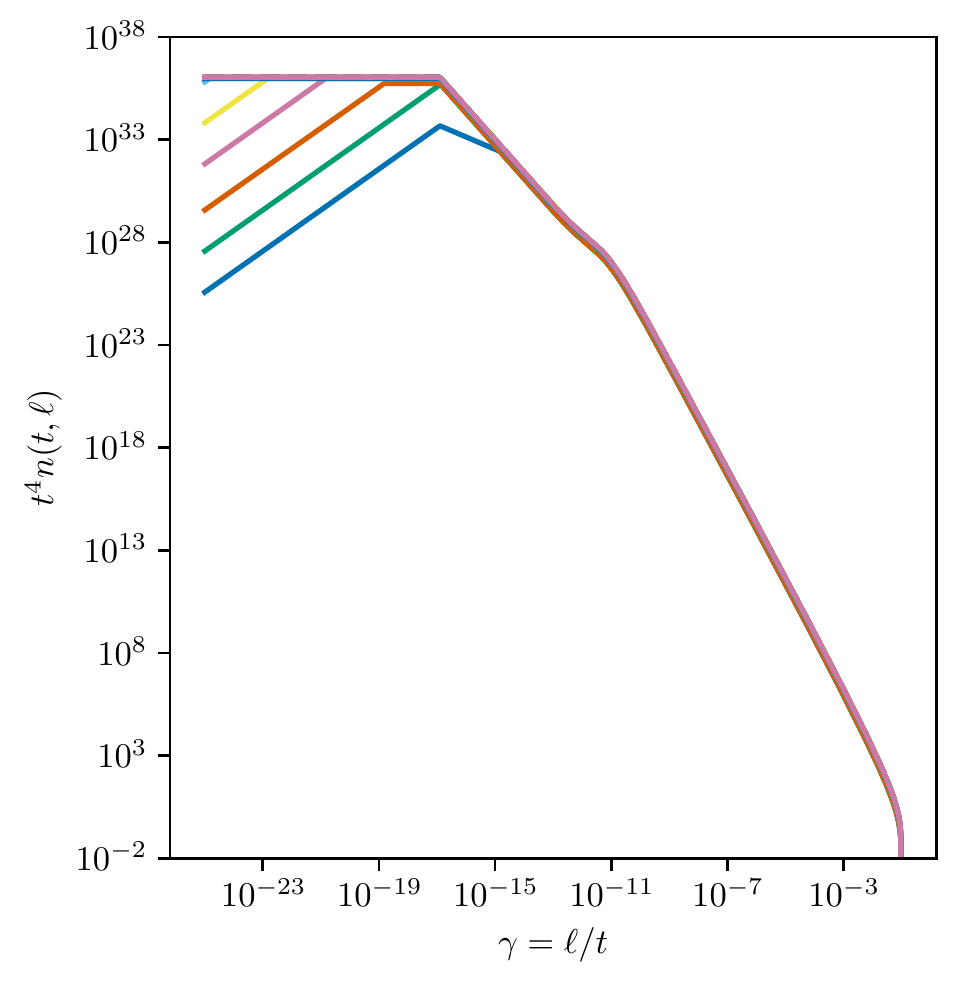}
	\includegraphics[width=.49\textwidth]{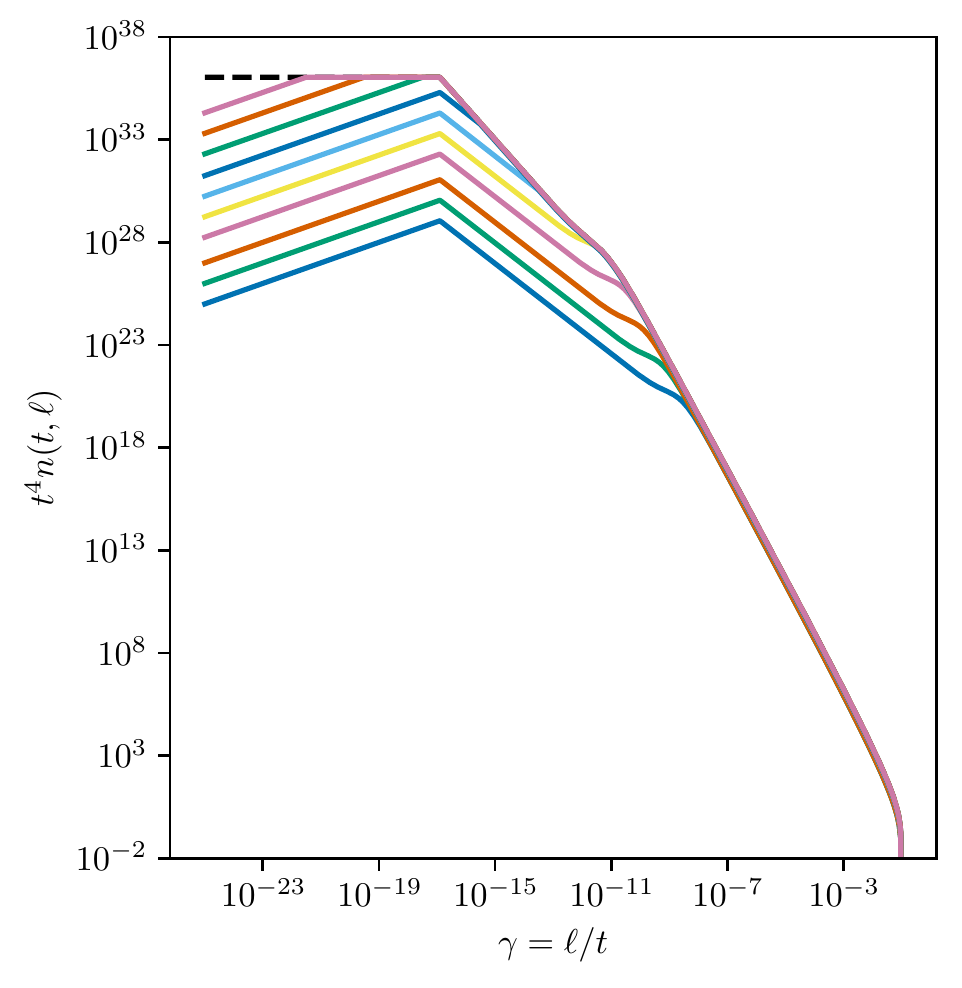}
	\caption{Loop number density for model B with kinks (left panel) and with cusps (right panel) for $G\mu = 10^{-13}$. Solid colored lines show the density for redshifts, from top to bottom $z = 10^6, 10^7, 10^8, 10^9, 10^{10}, 10^{11}, 10^{12}, 10^{13}, 10^{14}$ and $ 10^{15}$.
		The dashed dark line shows the scaling distribution when one assumes that all the energy goes into gravitational waves.}
	\label{fig:lnd}
\end{figure}

We now discuss the form of \cref{eq: final result n} in some extreme cases.
If $\ell \ll \ell_0$, then $\tau_0$ is very small and the second integral of the \cref{eq: final result n} vanishes.
Thus, the smallest loops are dissipated through particle emission, and the loop distribution is suppressed and non-scaling, see \cref{fig:lnd}.
In the opposite case when $\ell \gg \ell_0$, the first term vanishes and one recovers the scaling loop number density $n(t,\ell)$ with a constant loop decay, \emph{i.e.~}purely Nambu-Goto strings.
Physically, no particles are emitted in this latter case.
This corresponds to the regime $\gamma = \ell/t$ close to $1$ in \cref{fig:lnd}.

To obtain the total contribution of strings that are produced in matter and in the radiation dominated era, we can simply sum them as $n( t, \ell) = n_\mathrm{rad}( t, \ell) + n_\mathrm{mat}( t, \ell)\,.$

\begin{figure}
	\centering
	\includegraphics[width=.49\textwidth]{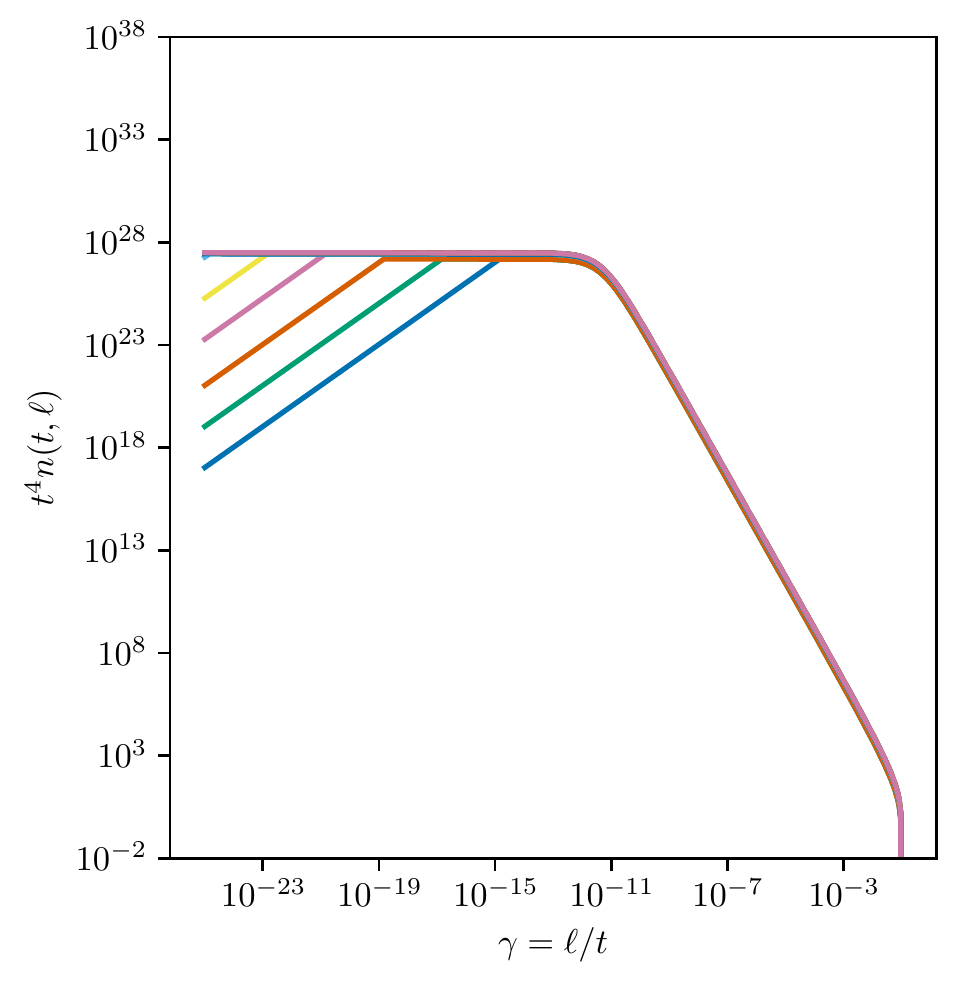}
	\includegraphics[width=.49\textwidth]{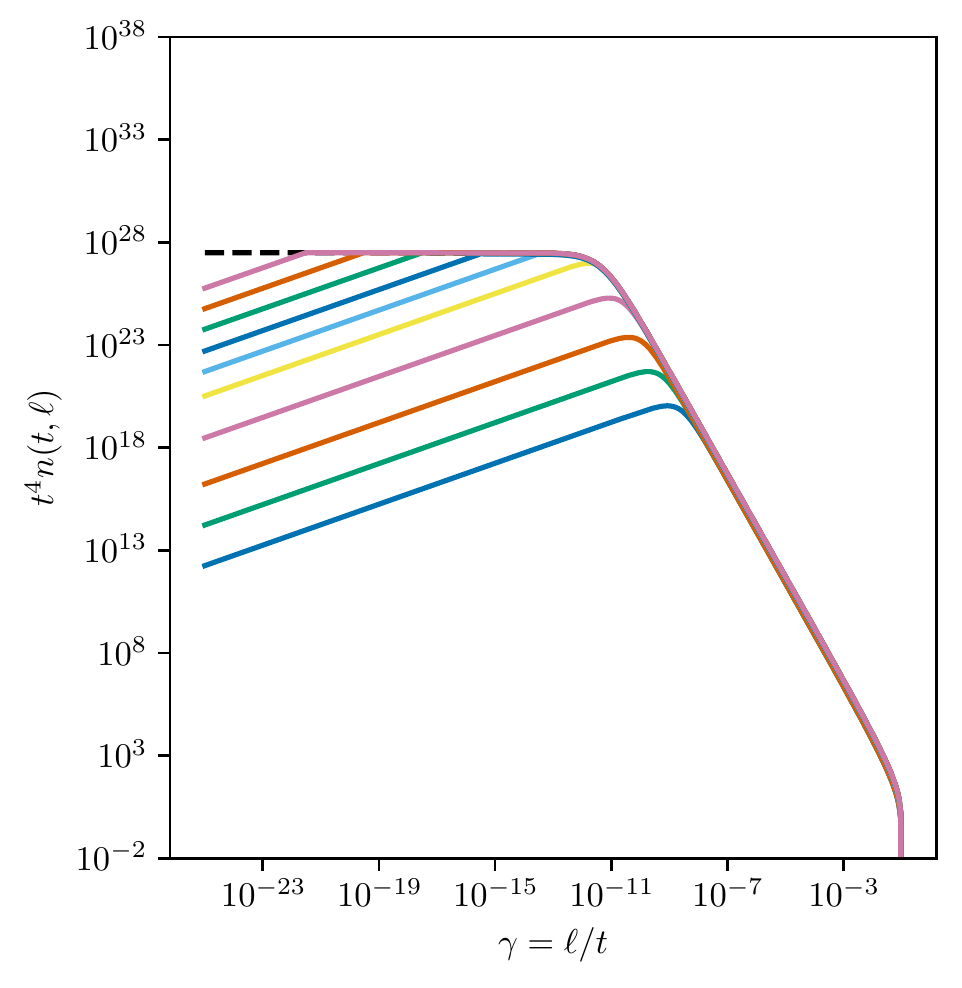}
	\caption{Loop number density for model A with kinks (left panel) and with cusps (right panel) for $G\mu = 10^{-13}$. Solid colored lines show the density for redshifts, from top to bottom $z = 10^6, 10^7, 10^8, 10^9, 10^{10}, 10^{11}, 10^{12}, 10^{13}, 10^{14}$ and $ 10^{15}$.
		The dashed dark line shows the scaling distribution when one assumes that all the energy goes into gravitational waves.}
\end{figure}

\section{Observational signatures}
\label{sec:obs}

From the loop distribution presented in \cref{sec:lnd}, we now calculate two observational signatures: the SGWB produced by cosmic string loops, and then the diffuse $\gamma$-ray background.
%
%In this section, we discuss two observational signatures for the loop distribution presented in \cref{sec:lnd}.
%First we calculate the resulting SGWB produced by cosmic string loops. Then, we constrain the cosmic string tension through their emission into particles.
%
Throughout we assume a standard Planck $\Lambda$CDM cosmology, with Hubble constant $\uHo=100 h \mathrm{km/s/Mpc}$, $h=0.678$, $\OmegaM=0.308$, $\OmegaR=9.1476\times 10^{-5}$ and $\Omega_{\Lambda} = 1-\OmegaM-\OmegaR$ \cite{Planck:2015fie}. The Hubble parameter $H(z) = \uHo \mathcal{H}(z)$ with
$\mathcal{H}(z) = \sqrt{\Omega_\Lambda + \OmegaM(1+z)^3 + \OmegaR \mathcal{{G}}(z)(1+z)^4 }$
where $\mathcal{G}(z)$ is directly related to the effective number of degrees of freedom $g_*(z)$ and the effective number of entropic degrees of freedom $g_S(z)$. More explicitly, \cite{Binetruy:2012ze}
\begin{equation}
	\mathcal{G}(z) = \frac{g_*(z) g_{S}^{4/3}(0)}{g_*(0) g_{S}^{4/3}(z)} =
	\begin{cases}
		\displaystyle
		1    & \text{for}\ z<10^9,              \\
		0.83 & \text{for}\
		10^9<z<2 \times 10^{12},                \\
		0.39 & \text{for} \ z>2 \times 10^{12}.
	\end{cases}
\end{equation}
which models the change in values at the QCD phase transition ($T=200$MeV), and at electron-positron annihilation ($T=200$keV).

\subsection{Stochastic background of gravitational waves}\label{sec:sgwb}

Individual cosmic string loops emit gravitational waves at a wide range of frequencies.
The power lost in GWs can be decomposed into a series of harmonics
\begin{equation}
	\Gamma G\mu^2 = G \mu^2 \sum_{j=1}^\infty P_j\,,
\end{equation}
where for small $j$, $P_j$ is determined by the oscillatory behavior of the loop. For $j \gg 1$ it is determined by the type of bursts (kinks or cusps) on the string~\cite{Vachaspati:1984gt,Burden:1985md,Garfinkle:1987yw,Hindmarsh:1990xi,Allen:1991bk}.
In particular, $P_j  \propto j^{-5/3}$ for kinks, whereas for loops with cusps $P_j  \propto j^{-4/3}$~\cite{Damour:2001bk,Siemens:2001dx,Siemens:2006vk,Blanco-Pillado:2017oxo}.
The incoherent contributions from all the loops across the history of the universe lead to a SGWB with spectrum \cite{Blanco-Pillado:2017oxo}
\begin{equation}
	\mathrm{\Omega}_\mathrm{gw}(\ln f) = \frac{8\pi (G\mu)^2 f}{3 \uHo^2}\sum_{j=1}^\infty C_j(f) P_j\,,
	\label{eqn:omega-method-1}
\end{equation}
where
\begin{equation}
	\label{eqn:Cn}
	C_j(f) = \frac{2j}{f^2 }\int_0^{\zf} \frac{\dd{z}}{H(z) (1+z)^6}~n\qty[t(z),\frac{2j} {(1+z)f}] \Theta\qty[\frac{2j} {(1+z)f} - \ell_0]\,,
\end{equation}
where we integrate from the friction dominated epoch $\zf$ until the present day.
Notice that we discard loops with sizes lower than $\ell_0$ since our assumption is that these decay into particles, see the energy budget of \cref{eq:loop decay}.

\begin{figure}
	\centering
	\includegraphics{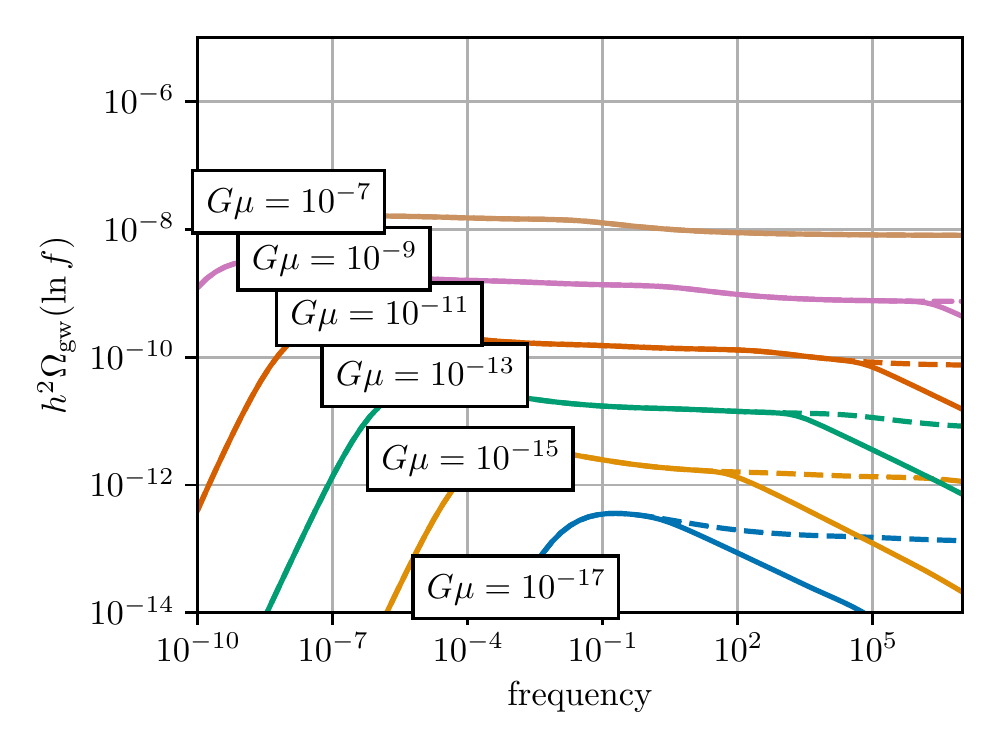}
	\includegraphics{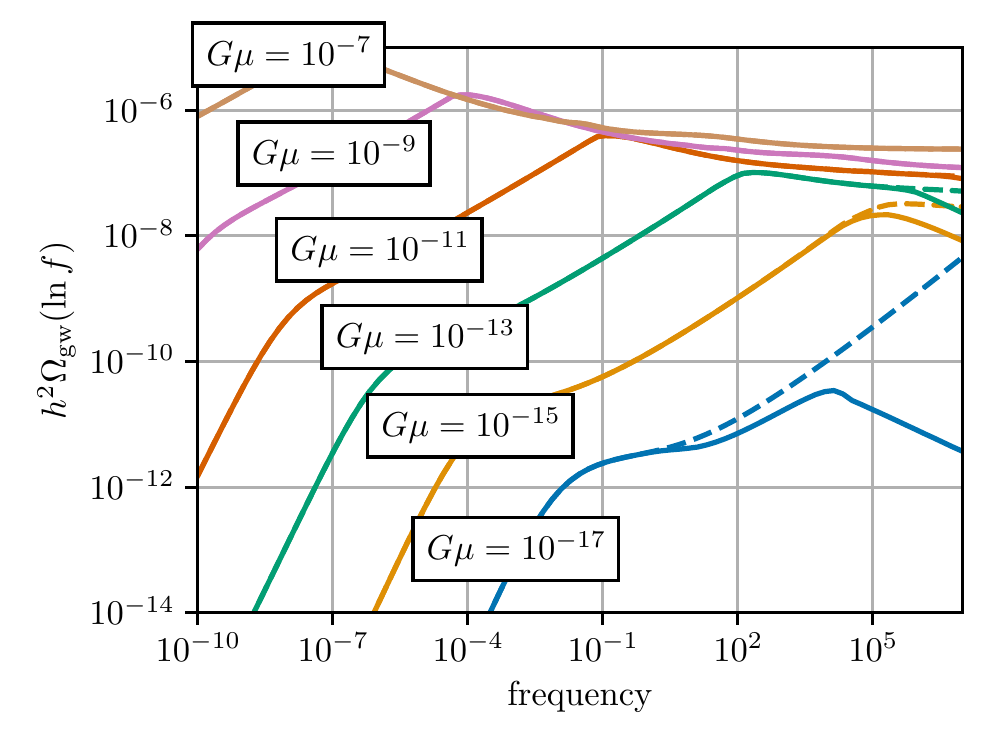}
	\caption{Stochastic background of gravitational waves for different $G\mu$.
	Top panel: Model A, lower panel: Model B.
	Solid line assume that cusps on the loops emit particles ($n = 1/2$). Dashed lines assume that all the emitted energy goes into gravitational waves.}
	\label{fig:sgwb}
\end{figure}

In \cref{fig:sgwb}, we present the GW power spectra in the presence of cusps, for $G\mu$ ranging from $10^{-7}$ to $10^{-17}$.
In Ref.~\cite{Auclair:2019jip}, the authors determined that, for a Dirac-delta loop production function (approximating Model A), the power spectrum was cutoff above a frequency
\begin{equation}
	f > \sqrt{\frac{8 \uHo \sqrt{\OmegaR}}{\ell_0 \Gamma G\mu}}.
\end{equation}
With the power-law loop production function of \cref{eq: def polchinski rocha production}, small loops of size $\ell / t \simeq \gbr$ mainly source the GW energy density.  We estimate that the GW spectra are cutoff above the frequency
\begin{equation}
	f \gtrsim \sqrt{\frac{8 \uHo \sqrt{\OmegaR}}{\ell_0 \gbr}}\, ,
\end{equation}
which, for Model B, is orders of magnitude higher than for Model A.
For kinks, the SGWB is modified at frequencies so large that it is indistinguishable from the standard NG scenario in all the frequency ranges available to present and planned GW experiments.
In \cref{fig:sgwb}, we observe that for cusps, the cutoff is well above the observable frequency range for ground-based detectors as long as $G\mu \gtrsim 10^{-17}$.
This means that the current bounds set by the LVK collaboration \cite{Abbott:2021ksc} are robust, even taking into account the emission of particles from cusps and kinks.

\subsection{Diffuse gamma ray background}\label{sec:gamma}

\begin{figure}
	\centering
	\includegraphics[width=0.49\textwidth]{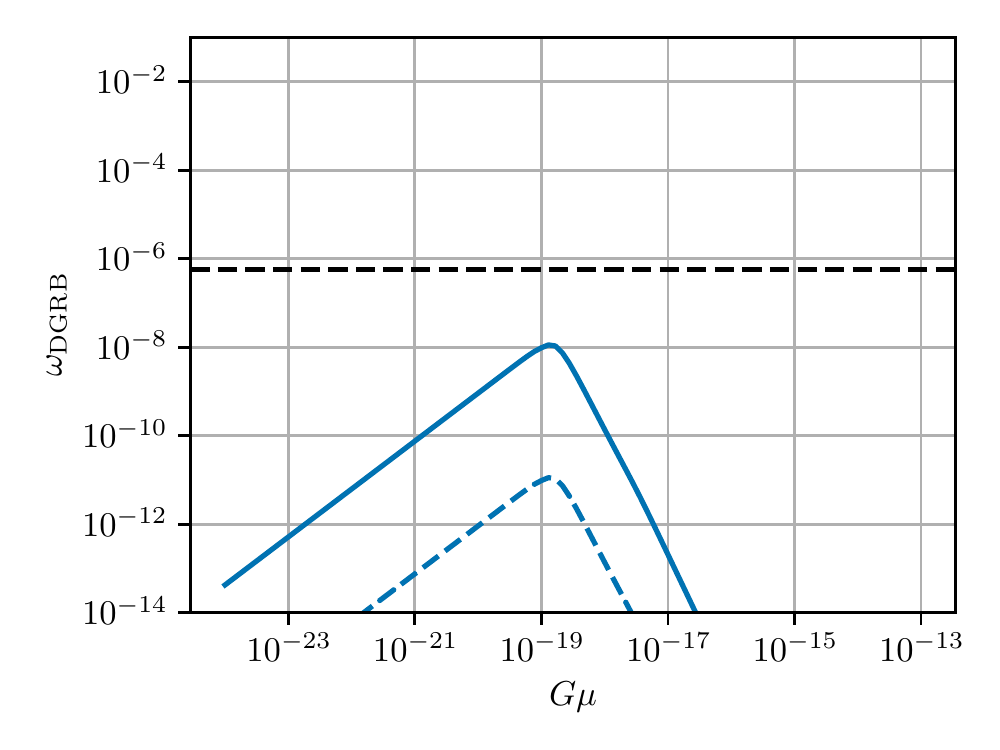}
	\includegraphics[width=0.49\textwidth]{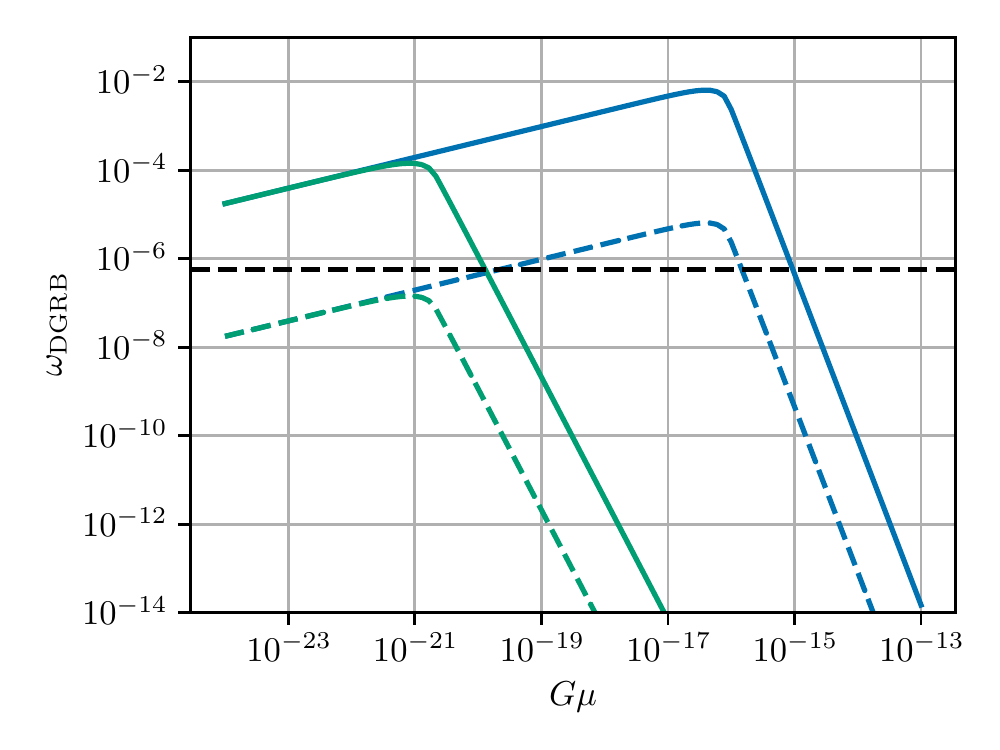}
	\caption{Diffuse $\gamma$-ray background in the presence of only cusps (blue) and only kinks (green) for $f_\mathrm{eff} = 1$ (solid lines) and $f_\mathrm{eff} = 10^{-3}$ (dashed lines).
	Left panel: Model A, right panel: Model B.}
	\label{fig:dgrb}
\end{figure}

In our model, loops decay into particles with power
\begin{equation}
	P_\mathrm{particles} = \Gamma G\mu^2 \qty(\frac{\ell_0}{\ell})^n, \quad \text{ for } \ell \leq \ell_0\,.
\end{equation}
Assuming that there is a coupling between the fields that make up the string and the Standard Model, this particle production will eventually cascade down into $\gamma$-rays with efficiency $f_\mathrm{eff} \leq 1$~\cite{Mota:2014uka}.
Thus, the string tension may be constrained by the DGRB measured by Fermi-LAT \cite{Fermi-LAT:2010pat}
\begin{equation}
	\odgrb^\mathrm{obs} \lesssim 5.8 \times 10^{-7} \mathrm{eV.cm}^{-3} ,
\end{equation}
where $\odgrb$ is the total energy density of GeV $\gamma$-rays injected since the universe became transparent at $t_\gamma \simeq 10^{15}$s~\cite{Mota:2014uka}.
The power injected to the dark sector at cosmic time $t$ is obtained by integrating the loop distribution over the loop sizes
\begin{equation}
	\Phi(t) = \Gamma G\mu^2 \int_0^{\gamma_\infty t} \calF \qty(\frac{\ell_0}{\ell})^n \Theta(\ell_0 - \ell)\dd{\ell}.
\end{equation}
The contribution of cosmic string loops to the DGRB today is therefore
\begin{align}
	\odgrb & = f_\mathrm{eff} \int_{t_\gamma}^{t_0} \frac{\Phi(t)\dd{t}}{[1+z(t)]^4}                                                                                                \\
	       & = f_\mathrm{eff} \Gamma G\mu^2 \ell_0^n \int_{t_\gamma}^{t_0} \frac{\dd{t}}{[1+z(t)]^4} \int_0^{\gamma_\infty t}  \frac{\calF}{\ell^n} \Theta(\ell_0 - \ell)\dd{\ell}.
	\label{eq:odgrb}
\end{align}
In \cref{fig:dgrb}, we show the expected DGRB expected from loops with only cusps (blue) and only kinks (green) as a function of the string tension $G\mu$ for $f_\mathrm{eff} = 1$.
For Model B, the power-law loop production function results in an abundant population of very small loops $\ell / t < \Gamma G \mu$, which enhances the DGRB and, as a result, excludes a large range of string tensions.
If we consider loops containing only cusps, the string tension is constrained to be $G\mu \gtrsim 10^{-15}$ for $f_\mathrm{eff} = 1$.
For loops containing only kinks the constraint is $G\mu \gtrsim 10^{-20}$.
Note that $\odgrb$ in \cref{eq:odgrb} depends linearly on $f_\mathrm{eff}$.
Referring to \cref{fig:dgrb}, this means that the string tension is constrained by the DGRB as long as $f_\mathrm{eff} > \order{10^{-3}}$.

The situation is very different with Model A for which the DGRB imposes no constraint.
It should be noted that this result confirms previous findings in Ref.~\cite{Auclair:2019jip} where the loop production function of Model A was approximated by a Dirac-Delta distribution.

\subsection{Joint constraints}
\label{sec:joint}

From observational constrains of GW interferometry, we find that the constraints on the string tension $G\mu \lesssim 4.0 \times 10^{-15}$ established by the LVK collaboration \cite{Abbott:2021ksc} remain valid when particle production is included.
Combining this with the lower constraint from the DGRB, the allowed region of parameter space of Model B is reduced to
\begin{equation}
	10^{-15}\lesssim (G\mu)_\text{cusps} \lesssim 4.0 \times 10^{-15}\,,
\end{equation}
in the case of cusps and
\begin{equation}
	10^{-20}\lesssim (G\mu)_\text{kinks} \lesssim 4.0 \times 10^{-15}\,,
\end{equation}
in the case of kinks for $f_\mathrm{eff} = 1$.
For cusps, the allowed window for $G\mu$ is very narrow and future experimental results from the LVK collaboration will reduce the upper bound; as a result in the coming years, either one will detect strings or rule out the existence of a string network having the properties assumed in this paper and \cref{tab:numirical values}.

In the coming decades, LISA is expected to probe the existence of cosmic strings with tension down to $G\mu \gtrsim 10^{-17}$~\cite{Auclair:2019wcv} limiting even more the available parameter space for $G\mu$.

\section{Conclusion}
\label{sec:conc}

In this work we have included the possible emission of both particles and gravitational waves into loop distribution models with the power-law loop production function of \cref{eq: def polchinski rocha production}.
For different choices of parameters, see \cref{tab:numirical values}, we recover Model A from Refs.~\cite{Blanco-Pillado:2013qja,Blanco-Pillado:2017oxo} and Model B from Refs.~\cite{Ringeval:2010ca,Lorenz:2010sm, Auclair:2019zoz}.
We find that the loop distribution is suppressed on small scales $\ell \ll \ell_0$ thus reducing the amount of energy emitted in GW at high frequencies, see also Ref.~\cite{Auclair:2019jip} which considered a Dirac-Delta loop production function.
With this power-law loop production function, loops are produced at all sizes at once, and the main contribution to the SGWB comes from smaller loops at a scale $\ell/t \lesssim \gbr$.
For Model A, we find similar constraints as in Ref.~\cite{Auclair:2019jip}.
For Model B, the SGWB is cut off at even larger frequencies and the current bound set by the third observing run of the LVK collaboration on this model, $G\mu \lesssim 4.0 \times 10^{-15}$, is robust.

We have computed the energy emitted into particles and calculated its contribution to the Diffuse $\gamma$-Ray Background.
We have assumed that the dark sector is coupled to the Standard Model and that the heavy particles emitted eventually decay into $\gamma$-rays with efficiency $f_\mathrm{eff} \leq 1$.
Under these assumptions, we find that, for Model B, string tensions of $G\mu \lesssim 10^{-20}$ are excluded in both the cusp-only and the kink-only scenarios.
In the worst case scenario that cusps dominate the radiation into particles and $f_\mathrm{eff} =1$, then the remaining window for the string tension is either narrowed down to $G\mu \approx 10^{-15}$ or completely closed.
Future observations by the LVK collaboration, and by LISA in the next decade will tighten the constraints on the string tension and Model B.

It remains to be determined what is the exact prevalence of cusps and kinks on cosmic string loops, but we expect the two limiting cases presented here to encompass most plausible results.
% Additionally, we should stress that the cosmic string model considered here, based on the Polchinski-Rocha loop production function, is only one amongst others and that there is not yet a consensus on the precise shape of the loop production function nor on the stability of loops (see for instance Ref.~\cite{Pazouli:2021orp} for recent work on loop fragmentation).
Finally, note that we have assumed that gravitational backreaction is the dominant mechanism for smoothing kinks on the infinite string network itself.
Particle production from kinks and kink-kinks collisions on infinite strings could also smooth them and prevent the formation of loops smaller than a certain scale, in a similar way.
This possibility deserves further analysis, and any particle production backreaction scale could be straightforwardly incorporated in \cref{eq: final result n}.

\section*{Acknowledgments}

We would like to thank Christophe Ringeval and Tanmay Vachaspati for useful discussions and encouragement. KL is grateful to the Fondation CFM pour la Recherche in France for support during his doctorate.
The work of PA is partially supported by the Wallonia-Brussels Federation Grant ARC \textnumero 19/24 - 103.

\appendix
\section{Models A and B: energy conservation}
\label{app:energy-con}

In this paper, we focus on Model B.
It has been argued in Ref. \cite{Blanco-Pillado:2019vcs} that this model does not respect energy conservation.
We now outline this argument and some of its limitations.
In particular we consider \emph{loop fragmentation}, namely the fact that loops formed from intercommutation are generally not stable, but fragment (on a given time-scale) after which they produce stable daughter loops. 
Loop fragmentation has been observed and studied in many simulations, including Refs.~\cite{Bennett:1985qt,Bennett:1986zn,Bennett:1987vf,Bennett:1989yp} as well as Refs.~\cite{Allen:1990mp,Allen:1990tv,Allen:1991jh}, and more recently Ref.~\cite{Ringeval:2005kr}.
In the simulations of Ref.~\cite{Blanco-Pillado:2019vcs}, on the contrary,   fragmentation is found to be quite rare.

We now recall briefly the argument of Ref.~\cite{Blanco-Pillado:2019vcs}, and then outline how it is modified by fragmentation (an effect which been included in other studies of loop evolution, see Ref. \cite{Copeland:1998na}).
%, which in particular has shown that it modifies the loop distribution at small scales leading to a much larger number of small loops (as one would expect),
The starting point of Ref.~\cite{Blanco-Pillado:2019vcs} is an equation for the energy density contained in the long string network,
\begin{equation}
\dv{\rho_\infty}{t} = -2 H (1 + \ev{v^2_\infty}) \rho_\infty - \mu \int_0^\infty \ell \calP(\ell, t) \dd{\ell}
\end{equation}
which, on assuming scaling, leads to
\begin{equation}
t^3 P_\mathrm{lost}(t) \equiv \int_0^\infty x \calP(x, t) \dd{x} = \frac{2}{\gamma^2}[1-\nu (1+\ev{v^2_\infty})] \equiv C
\label{eq:power}
\end{equation}
where $C$ is a constant whose value can be determined from the measured values of $\gamma$ and $\ev{v_\infty^2}$ in simulations, and  $P_\mathrm{lost}$ is the power lost into loop production by the infinite strings ($\calP(\ell, t)$ has been rescaled appropriately).
In Ref.~\cite{Blanco-Pillado:2019vcs}, the claim is that the loop production function of Model B gives an effective value of $C$ which is larger than that measured in their simulations by $\order{100}$.

How does loop fragmentation change this argument? The crucial point is that it is now no longer correct to equate the power \emph{lost} into loop production by the network $P_\mathrm{lost}(t)$ with the power \emph{received} by the \textbf{non self-intersecting} loops $P_\mathrm{rec}(t)$ \emph{at the same time}.

Indeed, due fragmentation, loops emitted from the infinite string network at time $t$ only form non self-intersecting loops at a later time $t'>t$ where, from \cref{eq:power},
\begin{equation}
\frac{P_\mathrm{lost}(t)}{P_\mathrm{lost}(t')} = \left(\frac{t'}{t}\right)^3 > 1.
\end{equation}
Thus the effective power going into the formation of \textbf{non self-intersecting} at time $t'$ is $P_\mathrm{lost}(t) > P_\mathrm{lost}(t')$.

We now estimate the size of this effect of loop fragmentation, under the standard assumptions that the network of infinite strings is scaling and loses its energy to loops of size $\alpha t_0$ at a time $t_0$.
We suppose\footnote{consistently with different simulation results} that these loops fragment after $\delta = \order{1}$ oscillations and, in order to keep the argument as simple as possible, we assume that the fragmentation is into two loops of identical size $\alpha t_0 / 2$.
Thus at $t_1 = (1 + \delta \alpha / 2 ) t_0$, all the loops formed at time $t_0$ fragment, and we can continue this cascade for an arbitrary number of fragmentations $n$
\begin{equation}
    t_n = t_{n-1} + \frac{\delta}{2} \frac{\alpha t_0}{2^{n-1}}  = \qty(1 + \delta\alpha \sum_{k=1}^{n} 2^{-k})t_0,
\end{equation}
after which we assume the loops are non self-intersecting.
Therefore $P_\mathrm{rec}(t_n) \dd{t_n} = P_\mathrm{lost}(t_0) \dd{t_0}$ and of course $P_\mathrm{rec}(t_n) \neq P_\mathrm{lost}(t_n)$.
Indeed, we can link the power lost by the network into loops at time $t_n$ with the power received by the non self-intersecting loops also at time $t_n$ using Eq. \eqref{eq:power}.
This results in a boost factor of
\begin{equation}
    \frac{P_\mathrm{rec}(t_n)}{P_\mathrm{lost} (t_n)} = \frac{P_\mathrm{lost}(t_0)}{P_\mathrm{lost} (t_n)} \dv{t_0}{t_n} = \qty(1 + \delta \alpha \sum_{k=1}^{n} 2^{-k})^2,
\end{equation}
accounting for the dilution of the power over a longer period of time.

For example, taking $\delta\alpha = 2$, then after five fragmentations the boost factor has nearly converged to $9$.
(These values are not unrealistic: the one-scale model predicts that the characteristic lengthscale of the network is $0.27 t$ during radiation era.
Therefore we take as fiducial values $\alpha = 0.27$ and $\delta \approx 6$.)
Hence this simple toy model shows that the ``energy conservation'' bound of Ref.~\cite{Blanco-Pillado:2019vcs} can be violated by a factor of $9$ by including a cascade of $5$ consecutive fragmentations into two equal parts.
Generally, however, the fragmentation will not be of loops of equal sizes, and the conclusion will be modified.
More accurate models have been discussed in the past such as in Ref.~\cite{Copeland:1998na}.
This illustrates the importance of fragmentation, among other physical effects, in accordance with energy balance.

\section{Calculation of integration bounds}
\label{app: calculation of tau technical}

In order to evaluate the bounds of \cref{eq: final result n} at the correct times, we have to calculate the start and end time of loop production, $\tstar$ and $\tbr$ respectively.
Recalling that $\tau = \Gamma G \mu t$ and using \cref{eq:v}, these satisfy
\begin{equation}
	\hat\xi(\gamma t') + \Gamma G \mu t' = 2 v
\end{equation}
where $2v = \hat\xi(\ell) + \hat \tau(t)$, which we need to solve for $t'$.
Because of the $\Theta$-function of \cref{eq: def xi}, $\hat \xi$ has two different forms which need to be considered separately.
If $\gamma t' > \ell_0$, then
\begin{equation}
	\gamma t' - n\frac{\ell_0}{(n+1)} + \Gamma G\mu t' = 2v \,,
	\label{eq:tprime-linear}
\end{equation}
which is straightforward to solve for $t'$.
If $\gamma t' < \ell_0$, then
\begin{equation}
	(\gamma t')^{n+1} - (n+1) \ell_0^n(2v - \Gamma G\mu t')= 0\,,
	\label{eq:tprime-poly}
\end{equation}
which is analytically soluble for $n \in \{ 0, 1, 1/2\}$.

\section{Energy lost by a cusp: an analytical and quantitative example}
\label{sec:cusp_analytic}

\begin{figure}
	\centering
	\includegraphics[trim=70 50 50 100, clip, width=.49\textwidth]{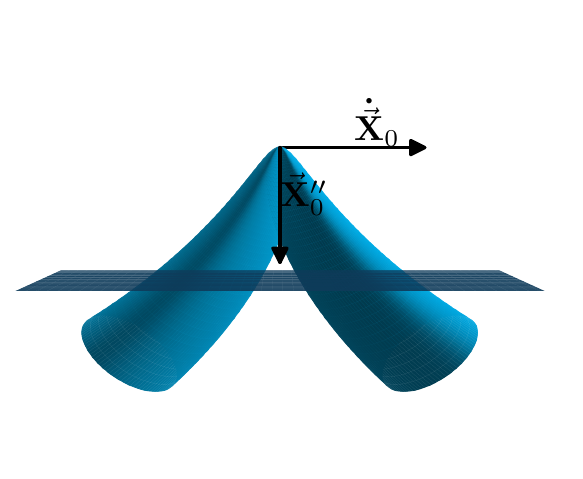}
	\includegraphics[width=.49\textwidth]{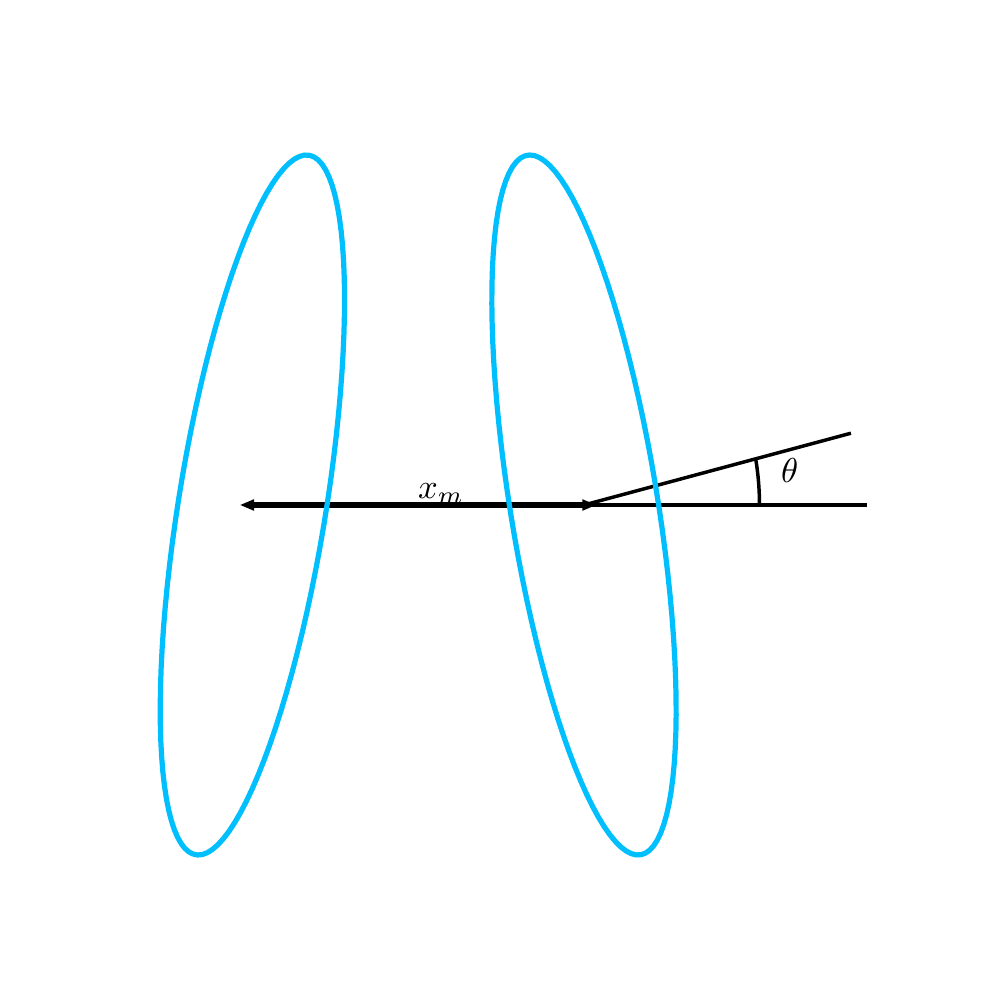}
	\caption{Left panel: One of the two cusps for a loop with $\alpha=3/4$ and an artificial width in the rest frame of the loop. The tip of the cusp goes at the speed of light in the direction of $\vb{\dot{X}}_0$. Right panel: a slice of the cusp along the plane parallel to $\vec{X}_0^{\prime\prime}$. The two branches of the cusp are separated by a distance $x_m$. The ellipses of the two branches are tilted with angle $\theta$.}
	\label{fig:cusp}
\end{figure}

In this appendix, we illustrate the argument first presented in \cite{Blanco-Pillado:1998tyu} concerning the energy lost into particles at a cusp. To do so, we work with the Kibble-Turok solution \cite{Kibble:1982cb} in Minkowski space-time.

The coordinates of the loop are
\begin{equation}
	X^\mu = X^\mu(\tau, \sigma),
\end{equation}
where $\tau$ and $\sigma$ are respectively time- and space-like coordinates on the loop worldsheet. Using the reparametrization invariance of the Nambu-Goto action, we fix the standard \emph{conformal-temporal gauge} in which $X^0 = \tau=t$. Then the spatial components $\vb{X}$ of the string satisfy
\begin{equation}
	\vb{X} = \frac{1}{2} \qty[ \vb{a}(\sigma - \tau) + \vb{b}(\sigma + \tau)],
\end{equation}
%where $\vb{X}$ is the three vector with spatial components.
%Additionally, our gauge choices impose that the solution satisfies the Virasoro constraints
where, from the gauge conditions,
\begin{align}
	\vb{X}^\prime \vdot \dot{\vb{X}}               & = 0  \\
	\norm{\vb{X}^\prime}^2 + \norm{\dot{\vb{X}}}^2 & = 1,
\end{align}
with $^\prime = \pdv*{\sigma}$ and $\dot{} = \pdv*{\tau}$.

In the center-of-mass frame, one of the simplest non-trivial examples for a loop of invariant size $\ell$, satisfying all these constraints, is the Kibble-Turok solution
\begin{equation}
	\vb{a^\prime}(u) =
	\mqty( (1-\alpha) \cos(\frac{2\pi u}{\ell}) + \alpha \cos(\frac{6\pi u}{\ell}) \\ (1-\alpha) \sin(\frac{2\pi u}{\ell}) + \alpha\sin(\frac{6\pi u}{\ell}) \\ 2 \sqrt{\alpha(1-\alpha)} \sin(\frac{2\pi u}{\ell}) ), \quad
	\vb{b^\prime} (v) = \mqty(\cos(\frac{2\pi v}{\ell}) \\ \sin(\frac{2\pi v}{\ell}) \\ 0)\,,
\end{equation}
where $u=\sigma-\tau$, $v=\sigma+\tau$, and $\alpha \in ]0, 1[$ is a free parameter that labels the different solutions in this family.
Each of these loops has exactly two cusps $\norm{\dot{\vb{X}}}^2 =1$ at $(\ell/4, \ell/4)$ and $(\ell/4, 3\ell/4)$, and we  analyze the former in the remainder of this appendix.
Furthermore, we assume that the portion of the string lost to particles is the overlap region around the cusp.
The distance between the two branches of the cusp can be found analytically
\begin{equation}
	x_m = \norm{\vb{X}(\ell/4, \ell/4 - \sigma) - \vb{X}(\ell/4, \ell/4+\sigma)}
	= \frac{2 \alpha \ell}{3\pi}\sin(\dfrac{2\pi \sigma}{\ell}).
\end{equation}
In the plane perpendicular to $\vb{X_0^{\prime\prime}}$ (see \cref{fig:cusp}), the section of the string is an ellipse with semi-major axis $w$, the width of the string, and semi-minor axis $\gamma w$, in which $\gamma$ is the Lorentz factor
\begin{equation}
	\gamma^{-1} = \sqrt{1 - \norm{\vb{\dot{X}}}^2} = \sqrt{\alpha} \sin(\dfrac{2\pi \sigma}{\ell}).
\end{equation}
Finally, the angle between the semi-minor axis and the two branches is
\begin{equation}
	\cos(\theta) = \frac{\sqrt{2}}{\sqrt{2 - \alpha + \alpha \cos(\frac{4\pi \sigma}{\ell})}}\cos(\dfrac{2\pi \sigma}{\ell}),
\end{equation}
thus the overlap region is determined by $\sigma_c$ satisfying the equality
\begin{equation}
	x_m^2(\sigma_c) = w^2 \left\{ \gamma(\sigma_c)^2 \cos^2[\theta(\sigma_c)] + \sin^2[\theta(\sigma_c)]\right\}.
\end{equation}
Finally, the energy of the string lost to particle by a single cusp is
\begin{equation}
	2 \mu \sigma_c = \frac{\mu \ell}{\pi} \arcsin(\sqrt{\frac{3\pi w}{\alpha \ell}}) = \mu \sqrt{\frac{3 w \ell}{\alpha \pi}} + \order{\frac{w}{\ell}}.
\end{equation}
Note that we obtained the same scaling $\sqrt{w\ell}$ as in Ref.~\cite{Blanco-Pillado:1998tyu}, and calculated analytically the prefactor for this family of loops.
The factor $1/\sqrt{\alpha}$ shows that the power emitted by different cusps may vary significantly, and we absorb this uncertainty with an effective factor $\beta$ in \cref{eq:ell0-kink,eq:ell0-cusp}.
Since a loop oscillates with period $\ell/2$, we estimate that the power emitted to particles due to one cusp is $\simeq \mu \sqrt{w}{\ell}$.

\bibliographystyle{JHEP}
\bibliography{refs}

\end{document}